\documentclass[longauth, traditabstract]{aa} 

\usepackage{graphicx}
\usepackage{txfonts}
\usepackage{natbib}
\begin{document}

\title{\emph{Herschel}/HIFI discovery of interstellar chloronium
  (H$_2$Cl$^+$)\thanks{\emph{Herschel} is an ESA space observatory with
    science instruments provided by European-led Principal
    Investigator consortia and with important participation from
    NASA.}}

\author{
D.~C.~Lis\inst{1},
J.~C.~Pearson\inst{13}, 
D.~A.~Neufeld\inst{3}, 
P.~Schilke\inst{8,12},  
H.~S.~P.~M\"{u}ller\inst{12},  
H.~Gupta\inst{13},
T.~A.~Bell\inst{1} , 
C.~Comito\inst{8}, 
T.~G.~Phillips\inst{1},
E.~A.~Bergin\inst{2}
% ...
}
\institute{California Institute of Technology, Cahill Center for Astronomy and Astrophysics 301-17, Pasadena, CA 91125 USA\\
             \email{dcl@caltech.edu}
\and Department of Astronomy, University of Michigan, 500 Church Street, Ann Arbor, MI 48109, USA 
\and  Department of Physics and Astronomy, Johns Hopkins University, 3400 North Charles Street, Baltimore, MD 21218, USA
\and Centre d'Etude Spatiale des Rayonnements, Universit\'e de Toulouse [UPS], 31062 Toulouse Cedex 9, France
\and CNRS/INSU, UMR 5187, 9 avenue du Colonel Roche, 31028 Toulouse Cedex 4, France
\and Laboratoire d'Astrophysique de l'Observatoire de Grenoble, 
BP 53, 38041 Grenoble, Cedex 9, France.
\and Centro de Astrobiolog\'ia (CSIC/INTA), Laboratiorio de Astrof\'isica Molecular, Ctra. de Torrej\'on a Ajalvir, km 4
28850, Torrej\'on de Ardoz, Madrid, Spain
\and Max-Planck-Institut f\"ur Radioastronomie, Auf dem H\"ugel 69, 53121 Bonn, Germany 
\and LERMA, CNRS UMR8112, Observatoire de Paris and \'Ecole Normale Sup\'erieure, 24 Rue Lhomond, 75231 Paris Cedex 05, France
\and LPMAA, UMR7092, Universit\'e Pierre et Marie Curie,  Paris, France
\and  LUTH, UMR8102, Observatoire de Paris, Meudon, France
\and I. Physikalisches Institut, Universit\"at zu K\"oln,
              Z\"ulpicher Str. 77, 50937 K\"oln, Germany
\and Jet Propulsion Laboratory,  Caltech, Pasadena, CA 91109, USA
% ...
}

\abstract{We report the first detection of chloronium, H$_2$Cl$^+$, in
  the interstellar medium, using the HIFI instrument aboard the
  \emph{Herschel} Space Observatory. The $2_{12}-1_{01}$ lines of
  ortho-H$_2^{35}$Cl$^+$ and ortho-H$_2^{37}$Cl$^+$ are detected in
  absorption towards NGC~6334I, and the $1_{11}-0_{00}$ transition of
  para-H$_2^{35}$Cl$^+$ is detected in absorption towards NGC~6334I
  and Sgr~B2(S). The H$_2$Cl$^+$ column densities are compared to
  those of the chemically-related species HCl. The derived
  HCl/H$_2$Cl$^+$ column density ratios, $\sim$1--10, are within the range
  predicted by models of diffuse and dense Photon Dominated Regions
  (PDRs). However, the observed H$_2$Cl$^+$ column densities, in
  excess of $10^{13}$~cm$^{-2}$, are significantly higher than the
  model predictions. Our observations demonstrate the outstanding
  spectroscopic capabilities of HIFI for detecting new interstellar
  molecules and providing key constraints for astrochemical models.}

   \titlerunning{Interstellar chloronium}
	\authorrunning{Lis et al.}
   \maketitle
%

\section{Introduction}

The halogen elements, fluorine and chlorine, form hydrides that are
very strongly bound: hydrogen fluoride is the only diatomic hydride,
and HCl$^+$ the only diatomic hydride cation, with a dissociation
energy exceeding that of molecular hydrogen. Drawing upon earlier work
by Jura (1974), Dalgarno et al. (1974), van~Dishoeck \& Black (1986),
Blake et al.\ (1986), Schilke et al.\ (1995), Federman et al.\ (1995),
and Amin (1996), Neufeld \& Wolfire (2009; hereafter NW09) have
recently carried out a theoretical study of the chemistry of
chlorine-bearing molecules, in both diffuse and dense molecular
clouds. In diffuse interstellar gas clouds, the dominant ionization
state of every element is determined by its ionization potential.
Chlorine, with an ionization potential slightly lower than that of
hydrogen, is predominantly singly-ionized. The Cl$^+$ ion can react
exothermically with H$_2$, the dominant molecular constituent of the
interstellar medium (ISM):
$$\rm Cl^+ + H_2 \rightarrow HCl^+ + H. \eqno{(1)}$$
The product of this reaction is the reactive HCl$^+$ ion, which
undergoes further reaction with H$_2$ to form H$_2$Cl$^+$:
$$\rm HCl^+ + H_2 \rightarrow H_2Cl^+ + H. \eqno{(2)}$$
The $\rm H_2Cl^+$ molecule does not react with H$_2$, and is destroyed
by dissociative recombination and proton transfer to CO, both of which
are sources of hydrogen chloride, HCl.

Prior to the launch of \emph{Herschel}, the H$^{35}$Cl and H$^{37}$Cl
isotopologues were the only chlorine-containing molecules to have been
detected in the ISM (e.g., Blake et al. 1985; Zmuidzinas et al. 1995;
Schilke et al. 1995; Salez et al. 1996; see also recent HIFI
observations of Cernicharo et al. 2010).\footnote{The metal halides
  NaCl, KCl, and AlCl have been detected in the {\it circumstellar
    envelope} of the evolved star IRC+10216, with abundances that
  reflect the thermochemical equilibrium established within the
  stellar photosphere (Cernicharo \& Gu\'elin 1987).} However,
predictions for the chemistry of Cl-bearing interstellar molecules
(NW09) have identified chloronium, $\rm H_2Cl^+$, as a relatively
abundant species that is potentially detectable. $\rm H_2Cl^+$ is
predicted to be most abundant in those environments where the
ultraviolet radiation is strong: in diffuse clouds, or near the
surfaces of dense clouds that are illuminated by nearby O and B stars.
In such environments, the photoionization of atomic chlorine leads to
a large abundance of Cl$^+$ ions that can form HCl$^+$ and $\rm
H_2Cl^+$ through reactions (1) and (2). A secondary abundance peak
occurs in dense, shielded regions; here HCl becomes a significant
reservoir of gas-phase chlorine, and can produce H$_2$Cl$^+$ through
reaction with $\rm H_3^+$:
$$\rm HCl + H_3^+ \rightarrow H_2Cl^+ + H_2. \eqno{(3)}$$
However, the chlorine depletion is typically large within such
regions (Schilke et al.\ 1995) and thus the overall $\rm H_2Cl^+$
abundance is rather small. 

In diffuse molecular clouds of density $n_{\rm H} = 10^{2.5} \rm \,
cm^{-3}$, H$_2$ column density $\ge 10^{20}\,\rm cm^{-2}$, and
$\chi_{\rm UV}$ in the range 1--10 (where $\chi_{\rm UV}$ is the UV
radiation field normalized with respect to the mean interstellar
value, Draine 1978), the NW09 model predicts $\rm H_2Cl^+$ column
densities $\sim$$3 \times 10^{10} \chi_{\rm UV} \,\rm cm^{-2}$. In
dense PDRs ($n_{\rm H} = 10^{4} \rm \, cm^{-3}$) illuminated by strong
radiation fields ($\chi_{\rm UV} > 10^3$), the predicted $\rm H_2Cl^+$
column densities are $\sim$10$^{12}$~cm$^{-2}$.

In this Letter, we report the first detection of chloronium towards
NGC~6334I and Sgr~B2(S), obtained using the HIFI instrument (de Graauw
et al. 2010) aboard the \emph{Herschel} Space Observatory (Pilbratt et al.
2010). NGC~6334 is a luminous and relatively nearby (1.7~kpc)
molecular cloud/H\,{\sc ii} region complex containing several
concentrations of massive stars at various stages of evolution. The
far-infrared source ``I'', located at the northeastern end of the
complex, is associated with a NIR cluster of bolometric luminosity of
$2.6 \times 10^5$ L$_\odot$ (Sandell 2000), with four embedded compact
millimeter continuum sources (Hunter et al. 2006). Sgr~B2(S) is a
strong submillimeter continuum source with a much less complex hot
core emission spectrum, as compared to its better known neighbor
Sgr~B2(M). This makes it a prime candidate for absorption studies,
probing the entire sight-line between the Sun and the Galactic center,
with clouds in the Orion, Sagittarius, and Scutum spiral arms easily
identified at separate velocities (e.g., Greaves \& Nyman 1996).

\section{Observations}

HIFI observations presented here were carried out between 2010 March 1
and March 23, using the dual beam switch (DBS) observing mode, as part
of guaranteed and open time key programs CHESS: Chemical
\emph{HErschel} Spectral Surveys, HEXOS: \emph{Herschel}/HIFI
Observations of EXtraOrdinary Sources: The Orion and Sagittarius B2
starforming regions, and HOP: \emph{Herschel} Oxygen Program. The
source coordinates are: $\alpha_{J2000} = 17^{\rm h}20^{\rm
  m}53.32^{\rm s}$ and $\delta_{J2000} =
-35^{\circ}46^{\prime}58.5^{\prime\prime}$ for NGC~6334I, and
$\alpha_{J2000} = 17^{\rm h}47^{\rm m}20.3^{\rm s}$ and
$\delta_{J2000} = -28^{\circ}23^{\prime}43.0^{\prime\prime}$ for
Sgr~B2(S).
The DBS reference beams lie approximately 3$^{\prime}$ east and west
(i.e. perpendicular to the roughly north--south elongation of the two
sources). Because the DBS mode alternates between two reference
positions, separated by 6$^{\prime}$ on the sky, we used the Level 1
data to compute a difference spectrum between the two reference
positions to check for possible contamination in the reference beams;
we see no evidence for emission or absorption in such a difference
spectrum. We used the HIFI wide band spectrometer (WBS) providing a
spectral resolution of 1.1~MHz ($\sim$0.4 km\,s$^{-1}$ at 780~GHz)
over a 4~GHz IF bandwidth. The spectra presented here are averages of
the H and V polarizations, with equal weighting, reduced using HIPE
(Ott 2010) with pipeline version 2.6. The resulting Level~2 DSB
spectra were exported to the FITS format for a subsequent data
reduction and analysis using the IRAM GILDAS package
(http://iram.fr/IRAMFR/GILDAS).

The band 2b, 1b and 1a spectral scans of NGC~6334I consist of double
sideband spectra (DSB) with a redundancy of 8, which gives
observations of a lower or upper sideband frequency with 8 different
settings of the local oscillator (LO). The Sgr~B2(S) data consist of 8
LO settings with a high redundancy of 12, centered near the frequency
of the 487.2~GHz line of O$_2$. The observations were fine-tuned so
that 4 of the 8 LO settings cover the frequency of the p-H$_2$Cl$^+$
line. These observing modes allow for the deconvolution and isolation
of a single sideband spectrum (Comito \& Schilke 2002). We applied the
standard deconvolution routine within CLASS. All NGC~6334I data
presented here are deconvolved single sideband spectra, including the
continuum. The HCl data in Sgr~B2(S) were obtained using the DBS
single point observing mode with 3 shifted LO settings that were
averaged to produce the final spectrum. The HIFI beam size at 485~GHz
and 780~GHz is 44$^{\prime\prime}$ and 30$^{\prime\prime}$,
respectively, with main beam efficiency of $\sim$0.68.

\section{Spectroscopy of H$_2$Cl$^+$}
\label{sec:disc}
The H$_{2}$Cl$^{+}$ ion is a closed-shell molecule, isoelectronic with
H$_{2}$S. Like H$_{2}$S and water, H$_{2}$Cl$^{+}$ is a highly
asymmetric top, exhibiting a {\it b}-type rotational spectrum. Its
fairly large dipole moment, calculated {\it ab initio} to be 1.89~D
(M{\"u}ller et. al 2005), about $70\%$ larger than that of HCl
(1.109~D; de Leluw \& Dymanus 1973), results in strong lines in the
THz range. Araki et al. (2001) measured rotational spectra of
H$^{35}_{2}$Cl$^{+}$, H$_{2}$$^{37}$Cl$^{+}$, and HDCl$^{+}$ below
500~GHz The accurate spectroscopic constants derived from these
measurements, including electric quadrupole coupling parameters, yield
a central bond angle in H$_{2}$Cl$^{+}$ of $\sim$94.2$^\circ$ (similar
to that of H$_{2}$S, $92.2^\circ$; Burrus \& Gordy 1953), and permit
the prediction of the ground state ortho transitions of
H$_{2}$Cl$^{+}$ and H$_{2}$$^{37}$Cl$^{+}$ near 780~GHz to well within
1~MHz (see Table~1 in the Online Material).

\section{Results}

\subsection{NGC~6334I}
Strong absorption at the frequency of the $2_{12}-1_{01}$ transition
of o-H$_2^{35}$Cl$^+$ at 781.6~GHz in NGC~6334I (Fig.~1a) has provided
the initial identification. Fitting the o-H$_2^{35}$Cl$^+$
$2_{12}-1_{01}$ hyperfine structure (HFS) gives a line velocity of
$-1.7$~km\,s$^{-1}$ for the strongest hyperfine component and a line
width of 11.6~km\,s$^{-1}$. The OH absorption profiles (Brooks \&
Whiteoak 2001) reveal two molecular clouds located along the line of
sight to NGC~6334, one with velocities extending from $-15$ to
2~km\,s$^{-1}$, and one with a well-defined velocity near
6~km\,s$^{-1}$. The hot core emission lines peak at about
$-6.5$~km\,s$^{-1}$ (e.g., C$^{18}$O 7--6 in the same band; HCl,
Sect. 4.1). Water and CH spectra towards NGC~6334I show multiple
velocity components, including absorption near 0~km\,s$^{-1}$, close
to the H$_2^{35}$Cl$^+$ velocity (Emprechtinger et al. 2010; van der
Wiel et al. 2010). The chloronium line velocity in NGC~6334I is in
good agreement with the H$_2$O$^+$ absorption velocity, also a tracer
of diffuse gas, when H$_2$O$^+$ frequencies of M\"{u}rtz et al. (1998)
are used (see Schilke et al. 2010 for a discussion of the H$_2$O$^+$
line frequencies). The large H$_2$Cl$^+$ line width may be due to
blending of multiple absorption components. However, the H$_2$O$^+$
line width is also quite large, about 8~km\, s$^{-1}$.

\begin{figure}[!htb]
   \centering
   \includegraphics[width=0.85\columnwidth]{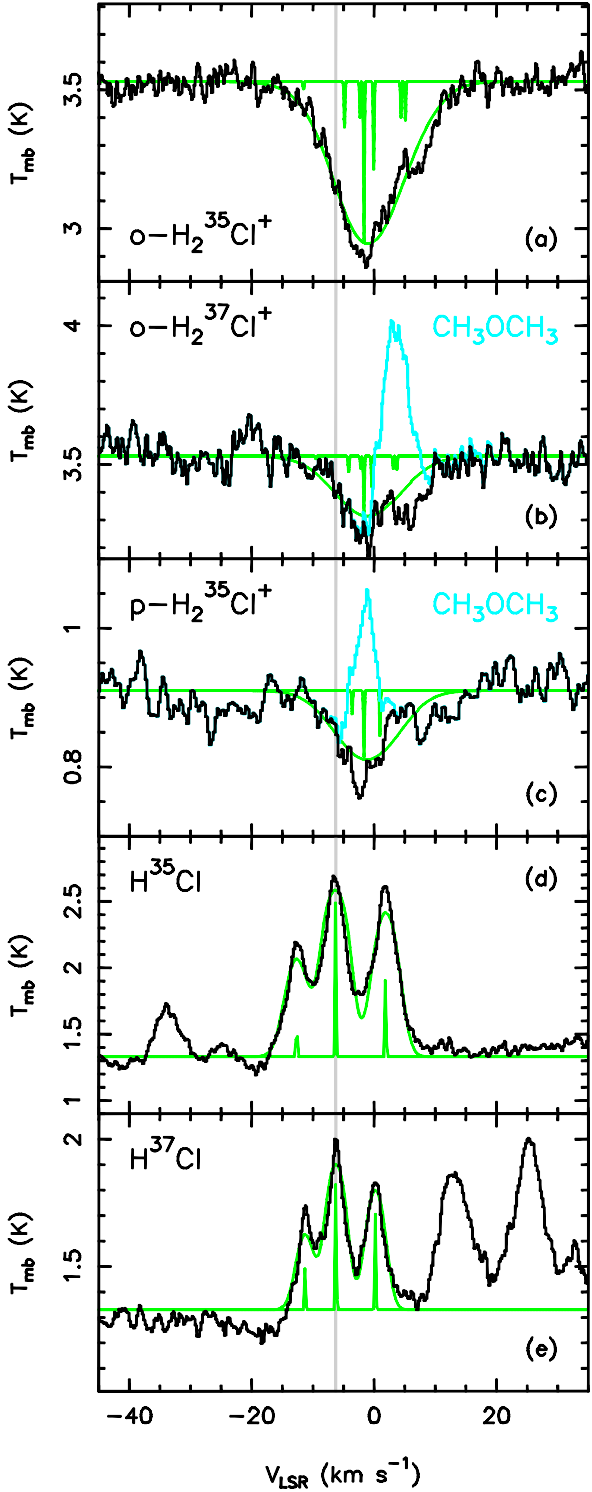}
  \caption{Spectra of chlorine species in NGC~6334I: (a)
     o-H$_2^{35}$Cl$^+$ $2_{12}-1_{01}$, (b) o-H$_2^{37}$Cl$^+$
     $2_{12}-1_{01}$, (c) p-H$_2^{35}$Cl$^+$ $1_{11}-0_{00}$, (d)
     H$^{35}$Cl $1-0$, and (e) H$^{37}$Cl $1-0$. The velocity scale
     corresponds to the strongest HFS components. Green lines show HFS
     fits and positions of the HFS components. The o-H$_2^{37}$Cl$^+$
     and p-H$_2^{35}$Cl$^+$ lines are blended with dimethyl ether
     emission (light-blue lines in panels b and c). }
         \label{fig1}
\end{figure}

The corresponding line of o-H$_2^{37}$Cl$^+$ is also detected
(Fig.~1b). However, the spectrum is contaminated by interfering
emission of dimethyl ether (light-blue line in Fig.~1b), one of
the most abundant ``weeds'' in NGC~6334I. The contamination is
subtracted by using an LTE model that fits profiles of nearby dimethyl
ether lines with similar upper level energies (Endres et al. 2009).
The resulting o-H$_2^{37}$Cl$^+$ spectrum is shown as a black line in
Fig.~1b. The $1_{11}-0_{00}$ line of p-H$_2^{35}$Cl$^+$ (Fig.~1c) is also
blended with dimethyl ether emission, similarly subtracted.

H$_2$Cl$^+$ spectra in NGC~6334I can be compared to those of the
chemically related species HCl (Fig.~1d and 1e). The lines of
H$^{35}$Cl and H$^{37}$Cl are detected \emph{in emission} at the hot
core velocity ($\sim$$-6.3$~km\,s$^{-1}$), with narrow line widths of
4.1 and 3.3 km\,s$^{-1}$, respectively. The HCl HFS is resolved
spectrally using HIFI WBS, allowing for the determination of the line
opacity.

We have modelled the H$_2$Cl$^+$ spectra assuming the same
excitation temperature of 5~K for all hyperfine components.\footnote{We made
  use of the myXCLASS program
  (http://www.astro.uni-koeln.de/projects/schilke/XCLASS), which
  accesses the CDMS (M\"uller at al. 2001, M\"uller et al. 2005;
  http://www.cdms.de) and JPL (Pickett et al. 1998;
  http://spec.jpl.nasa.gov) molecular databases.} 
A low value of the excitation temperature is justified given the high
spontaneous emisison rates and critical densities of the transitions
considered here and it provides a lower limit for the molecular column
densities derived from absorption measurements. An HFS fit to the
$2_{12}-1_{01}$ transition of o-H$_2^{35}$Cl$^+$ (green line in
Fig.~1a) gives an o-H$_2^{35}$Cl$^+$ column density of $1.3 \times
10^{13}$~cm$^{-2}$, under the assumption that the absorption
completely covers the continuum and is not concentrated in small
clumps. A fit to the $1_{11}-0_{00}$ spectrum of p-H$_2^{35}$Cl$^+$,
with all parameters other than the column density fixed, gives a
p-H$_2^{35}$Cl$^+$ column density of $4.0 \times 10^{12}$~cm$^{-2}$.
The total H$_2^{35}$Cl$^+$ column density is thus $1.7 \times
10^{13}$~cm$^{-2}$ and the ortho-to-para ratio is 3.2, consistent with
the statistical weight ratio. For an excitation temperature of 2.7~K,
the ortho and para H$_2$Cl$^+$ column densities are approximately 10\%
and 20\% lower, respectively. The H$_2^{35}$Cl$^+$ spectra are all
optically thin (line center optical depth of $\sim$0.2 for the ortho
line). We derive an H$_2^{35}$Cl$^+$/H$_2^{37}$Cl$^+$ ratio of 3,
close to the terrestrial ratio of 3.1.

We have modelled the H$^{35}$Cl and H$^{37}$Cl emission spectra
assuming a source size of 10$^{\prime\prime}$ (approximate size of the
cluster of compact continuum sources seen in the SMA image of Hunter
et al. 2006). Under this assumption, a least squares fit to the
H$^{35}$Cl spectrum gives an excitation temperature of 31~K and a
column density of
$4.0 \times 10^{14}$~cm$^{-2}$. For H$^{37}$Cl, we derive an
excitation temperature of 21~K and a column density of
$1.5 \times 10^{14}$~cm$^{-2}$; the resulting H$^{35}$Cl/H$^{37}$Cl
ratio is 2.7. However, HCl column densities and the isotopic ratio
depend strongly on the assumed source size (for a source size of
5$^{\prime\prime}$ the derived isotopic ratio is 4.1). The 350 $\mu$m
continuum flux density toward NGC~6334I is 1430~Jy in a
9$^{\prime\prime}$ beam (CSO/SHARC~II; Dowell et al., private comm.)
Assuming a dust temperature of 100~K (Sandell 2000) and a grain
emissivity $\kappa_{350} = 0.1$~cm$^{-2}$g$^{-1}$, we derive an H$_2$
column density of $1.2 \times 10^{24}$~cm$^{-2}$, which implies an
H$^{35}$Cl abundance of $\sim$$1.7 \times 10^{-10}$ with respect to H
nuclei. The lines of both HCl isotopologues are optically thick, with
line center optical depths of $\sim$2.2 and 1.6 for the strongest
hyperfine component of H$^{35}$Cl and H$^{37}$Cl, respectively.

\subsection{Sgr~B2(S)}

The p-H$_2^{35}$Cl$^+$ spectrum towards Sgr~B2(S) (Fig.~2, upper
panel) shows strong absorption near the systemic velocity of the
Sgr~B2 envelope ($\sim$62~km\,s$^{-1}$) and two additional deep
absorption components between 0 and 20~km\,s$^{-1}$. In addition,
shallow absorption is seen over a broad range of velocities down to
$-100$~km\,s$^{-1}$, in agreement with the H\,{\sc i} absorption
spectrum towards the nearby source Sgr~B2(M) (magenta line in Fig.~2).
Both H$^{35}$Cl and H$^{37}$Cl (Fig.~2, lower panel) show deep
absorption at the envelope velocity and a shallow absorption between 0
and 20~km\,s$^{-1}$. Similarly to NGC~6334I, we see velocity offsets
of order a few km\,s$^{-1}$ between HCl and H$_2$Cl$^+$ components.

\begin{figure}
   \centering
   \includegraphics[width=0.92\columnwidth]{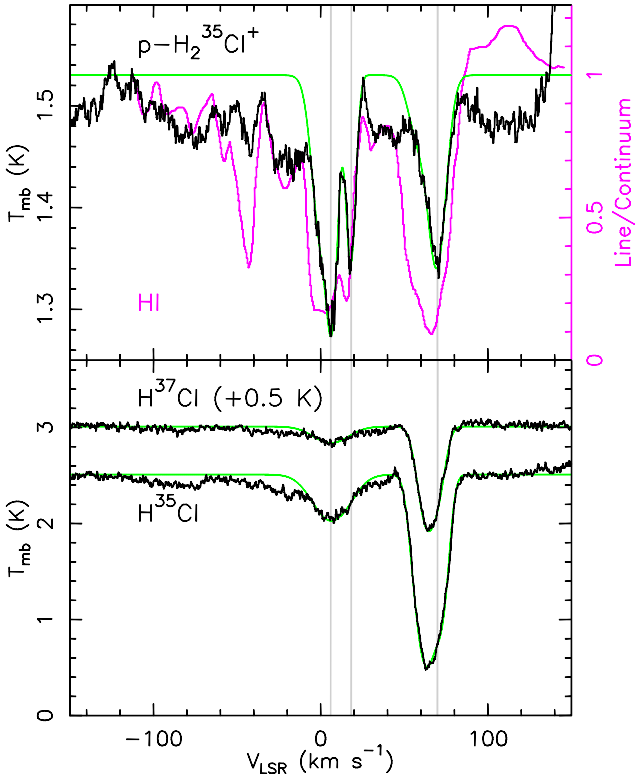}
   \caption{Spectra of p-H$_2^{35}$Cl$^+$$1_{11}-0_{00}$ (upper
     panel), and H$^{35}$Cl and H$^{37}$Cl $1-0$ (lower panel) towards
     Sgr B2(S). The H$^{37}$Cl spectrum has been shifted up by 0.5~K.
     The magenta line in the upper panel shows the H\,{\sc i}
     absorption spectrum towards Sgr~B2(M) (from Garwood \& Dickey
     1989).}
         \label{fig2}
\end{figure}

Assuming a 5~K excitation temperature (the same as for NGC~6334I) and
an ortho/para ratio of 3, we derive H$_2$Cl$^+$ column densities of
$3.4 \times 10^{13}$ and $2.2 \times 10^{13}$~cm$^{-2}$ for the 0 and
62~km\,s$^{-1}$ components, with corresponding H$^{35}$Cl column
densities of
$4 \times 10^{13}$ and $2 \times 10^{14}$~cm$^{2}$. The
H$^{35}$Cl/H$^{37}$Cl ratio is $\sim$3.3 in both components. We
estimate the uncertainties in our molecular column density estimates to be of
order a factor of 2.

To derive the hydrogen column density in the foreground gas towards
Sgr~B2(S), we use the method employed in Lis et al. (2001) to
analyze the O\,{\sc i} absorption towards Sgr~B2(M), based on
H\,{\sc i} and $^{13}$CO absorption data. We assume that the
foreground absorption is extended and column densities are the same
towards Sgr~B2(M) and (S). We derive a total hydrogen nuclei column
density of $\sim$$2 \times 10^{22}$~cm$^{-2}$ in the atomic and
molecular components in the velocity range $-10$ to 20~km\,s$^{-1}$
(with a factor of 2 uncertainty). The corresponding chlorine content,
in the form of H$_2$Cl$^+$ and HCl, is $7 \times
10^{13}$~cm$^{-2}$, implying a Cl/H ratio of $\sim$$4 \times 10^{-9}$.
This can be compared to the values measured in the UV in diffuse
clouds (e.g., Sonnentrucker et al. 2006), which are in the range $3
\times 10^{-8} - 4 \times 10^{-7}$. Therefore the high H$_2$Cl$^+$
column densities we derive here are consistent with the overall
chlorine budget, leaving plenty of room for atomic Cl and depletion on
dust grains. 

\section{Discussion}

Our estimates of the H$_2$Cl$^+$ column densities towards NGC~6334I
and Sgr~B2(S), in excess of $10^{13}$~cm$^{-2}$, are significantly
higher than those expected for a single dense or diffuse PDR
viewed at normal incidence. This might point to some deficiency in the
models. Alternatively, a significant enhancement in the absorbing
column density could result if the normal to the irradiated surface
were inclined relative to the sight-line, or indeed if multiple PDRs
were present 
along the sight-line, particularly if the radiation field is
enhanced, as may be likely for the multiple absorption components seen
towards Sgr~B2. Similar discrepancies between models and observations
are seen for other reactive ions in massive starforming regions (e.g.,
CO$^+$ toward AFGL 2591; Bruderer et al. 2009).

We derive an HCl/H$_2$Cl$^+$ ratio of $\sim$10 in NGC~6334I and the
Sgr~B2 envelope (assuming that in the case of NGC~6334I the
H$_2$Cl$^+$ column density on the back side is the same as that
derived in front of the continuum source from our absorption
measurements). This is well within the range predicted for dense PDRs
(up to $\sim$100 for densities above $10^6$~cm$^{-3}$). The
HCl/H$_2$Cl$^+$ ratio derived in the foreground gas towards Sgr~B2(S)
at velocities $0-20$~km\,s$^{-1}$, $\sim$1, is also consistent with
predictions of diffuse cloud models.

While a detailed analysis of chlorine chemistry in these and other
sources that have been or will be observed using HIFI will be
presented in a forthcoming paper, this work clearly demonstrates the
outstanding spectroscopic capabilities of HIFI in the search for new
interstellar molecules, particularly hydrides, and in providing robust
constraints for astrochemical models of the interstellar medium.

\onltab{1}{
\begin{table}
\label{tab:spec}
\caption{Frequencies of the H$_2$Cl$^+$ transitions observed.}
\begin{tabular}{rcccc}
\hline \hline
\rule[-3mm]{0mm}{8mm}$Transition$ & $Frequency$ & $Error$ & $A_{ij}$  & $E_l$\\
&  ( MHz)      & (MHz)    &              (s$^{-1}$)     & (cm$^{-1}$)  \\
\hline
\multicolumn{5}{c}{\rule[-3mm]{0mm}{8mm}H$_2^{35}$Cl$^+$} \\
$1_{11}-0_{00}~3/2-3/2$   &  485413.427 &  0.029  & 0.00159 & 0 \\
$5/2-3/2$                        &  485417.670 &  0.015  & 0.00159 & 0 \\
$1/2-3/2$                        &  485420.796 &  0.057  & 0.00159 & 0\\

$2_{12}-1_{01}~5/2-5/2$    & 781609.303 & 0.063 & 0.00179 & 14.1\\
$3/2-1/2$                        & 781611.062  & 0.062 & 0.00248 &  14.1\\
$5/2-3/2$                        & 781622.721  & 0.063 & 0.00417 &  14.1\\
$7/2-5/2$                        & 781626.794  & 0.060 & 0.00596 &  14.1\\
$1/2-1/2$  &781628.554  & 0.061 & 0.00496 &  14.1\\
$3/2-3/2$  &781635.214  & 0.062 & 0.00318 &  14.1\\

\hline
\multicolumn{5}{c}{\rule[-3mm]{0mm}{8mm}H$_2^{37}$Cl$^+$} \\

$2_{12}-1_{01}~5/2-5/2$  & 780037.315  & 0.069 & 0.00178 &  14.1\\
$3/2-1/2$ &  780038.760  & 0.066 & 0.00247 &   14.1 \\
$5/2-3/2$ &  780047.903  & 0.066 & 0.00414 &   14.1 \\
$7/2-5/2$ &  780051.197  & 0.062 & 0.00592 &   14.1 \\
$1/2-1/2$ &  780052.642  & 0.065 & 0.00493 &   14.1 \\
$3/2-3/2$ &  780057.820  & 0.068 & 0.00316 &   14.1 \\
\hline
\end{tabular}
Note: Frequencies and spontaneous emission coefficients have been
calculated from the constants derived by Araki et
al. (2001), see also CDMS. 
\end{table}

\begin{acknowledgements}
  HIFI has been designed and built by a consortium of institutes and
  university departments from across Europe, Canada and the United
  States under the leadership of SRON Netherlands Institute for Space
  Research, Groningen, The Netherlands and with major contributions
  from Germany, France and the US. Consortium members are: Canada:
  CSA, U.Waterloo; France: CESR, LAB, LERMA, IRAM; Germany: KOSMA,
  MPIfR, MPS; Ireland, NUI Maynooth; Italy: ASI, IFSI-INAF,
  Osservatorio Astrofisico di Arcetri- INAF; Netherlands: SRON, TUD;
  Poland: CAMK, CBK; Spain: Observatorio Astronomico Nacional (IGN),
  Centro de Astrobiolog\'ia (CSIC-INTA). Sweden: Chalmers University
  of Technology - MC2, RSS \& GARD; Onsala Space Observatory; Swedish
  National Space Board, Stockholm University - Stockholm Observatory;
  Switzerland: ETH Zurich, FHNW; USA: Caltech, JPL, NHSC. Support for
  this work was provided by NASA through an award issued by
  JPL/Caltech. D.~C.~L. is supported by the NSF, award AST-0540882 to
  the CSO. A portion of this research was performed at the Jet
  Propulsion Laboratory, California Institute of Technology, under
  contract with the National Aeronautics and Space Administration.
\end{acknowledgements}
}

\end{document}